\documentclass[aip,apl,graphicx,reprint]{revtex4-1}
\usepackage{graphicx}

\begin{document}


\title{Perpendicular magnetic anisotropy in Co$_2$Fe$_{0.4}$Mn$_{0.6}$Si} 



\author{B.\,M.\,Ludbrook}
\author{B.\,J.\,Ruck}

\affiliation{The MacDiarmid Institute for Advanced Materials and Nanotechnology, School of Chemical and Physical Sciences, Victoria University of Wellington, P.O. Box 600, Wellington 6140, New Zealand}

\author{S.\,Granville}
\affiliation{The MacDiarmid Institute for Advanced Materials and Nanotechnology, Robinson Research Institute, Victoria University of Wellington, P.O. Box 33436, Lower Hutt 5046, New Zealand}

\date{\today}

\begin{abstract}
We report perpendicular magnetic anisotropy (PMA) in the half-metallic ferromagnetic Heusler alloy Co$_2$Fe$_{0.4}$Mn$_{0.6}$Si (CFMS) in a MgO/CFMS/Pd trilayer stack. PMA is found for CFMS thicknesses between 1 and 2~ nm, with a magnetic anisotropy energy density of $K_U = 1.5\times 10^6$~erg/cm$^3$ for t$_{\tiny \textrm{CFMS}} = 1.5$~nm. Both the MgO and Pd layer are necessary to induce the PMA. We measure a tunable anomalous Hall effect, where its sign and magnitude vary with both the CFMS and Pd thickness.

\end{abstract}


\maketitle 


A magnetic thin film will generally prefer to have its magnetic moment lying in the plane of the sample owing to the large demagnetizing field. New magnetic devices have made it desirable to fabricate thin magnetic layers which instead have an easy axis of magnetization directed perpendicular to the film plane, a condition known as perpendicular magnetic anisotropy (PMA).\cite{Hirohata2015} In particular, spin-transfer-torque (STT) based devices require that a magnetic free layer can be easily flipped by a spin-polarized current while maintaining a high stability against thermal fluctuations.\cite{Mangin2006} Magnetic layers with PMA are well suited to optimize this trade-off for device applications. 

In order to realize a high efficiency STT device, a high degree of spin-polarization in the magnetic layers is also desirable. CoFeB, with approximately 65\% spin polarization, has been the most widely studied material so far, because it can be grown with PMA and incorporated into device structures with very large tunneling magnetoresistance (TMR).\cite{Ikeda2010} There is a strong motivation to incorporate a half-metallic ferromagnet (HMFM) with higher spin polarization in these devices, and Heusler alloys are promising in this regard.\cite{Katsnelson2008} In particular, Co$_2$FeSi and Co$_2$MnSi both have 100\% spin polarization\cite{Kandpal2006,Yang2013,Jourdan2014} and high Curie temperatures (c. 1000$^{\circ}$C). Recent studies have shown the intermediary compound Co$_2$Fe$_x$Mn$_{(1-x)}$Si with x$\approx$0.4 to be eminently promising for device applications with a low Gilbert damping parameter\cite{Kubota2009} and record 75\% room temperature GMR ratio.\cite{Sato2011}

Efforts to induce PMA in the Heuslers have focused on compounds containing Fe on MgO (eg. Co$_2$FeAl\cite{Wen2011,Li2011a}). This was guided by earlier studies of CoFeB/MgO where the PMA is thought to have its origin in the Fe-O hybridization.\cite{Ikeda2010} PMA has recently been reported in Co$_2$MnSi in CMS/Pd multilayer stacks on MgO,\cite{Matsushita2015} and for Pd buffered Co$_2$Fe$_{x}$Mn$_{1-x}$Si on MgO,\cite{Kamada2014} but the details of the PMA and the contribution of the various interfaces remains unclear. Here, we demonstrate PMA in MgO/CFMS/Pd stacks, and show that both interfaces are important for this effect.

Samples were grown by DC magnetron sputtering in a Kurt J Lesker CMS-18 UHV system with a base pressure of $2\times 10^{-8}$ Torr. Multilayer stacks consisting of MgO(2)/CFMS(t$_{\tiny\textrm{CFMS}}$)/Pd(2.5), where the number in parentheses is the nominal layer thickness in nm, were grown on $10 \times 10$mm Si/SiO$_2$ substrates at room temperature and post-growth annealed in-situ for 1 hour at 300$^{\circ}$C with an in-plane magnetic field of 170 Oe. MgO was RF sputtered with 100 W power in 3 mTorr Ar, giving a growth rate of 0.05 \AA/s. CFMS was DC sputtered at 100 W and 5 mTorr Ar, giving a growth rate of 0.43 \AA/s, and Pd was DC sputtered at 175 W in 8 mTorr Ar, with a growth rate of 4.0 \AA/s. Growth rates were calculated by growing a thick ($>50$~nm) film and measuring the thickness with a Dektak profilometer. The composition of the Co$_2$Fe$_{0.4}$Mn$_{0.6}$Si target was verified by energy dispersive x-ray analysis in a SEM. Magnetization and Hall resistance measurements were done in a Quantum Design SQUID and PPMS respectively, at room temperature. The Hall measurements were made using sampleholders from Wimbush Science \& Technology, with spring-loaded contacts in a van der Pauw geometry.


\begin{figure}[h!]
\includegraphics[width=8.5cm]{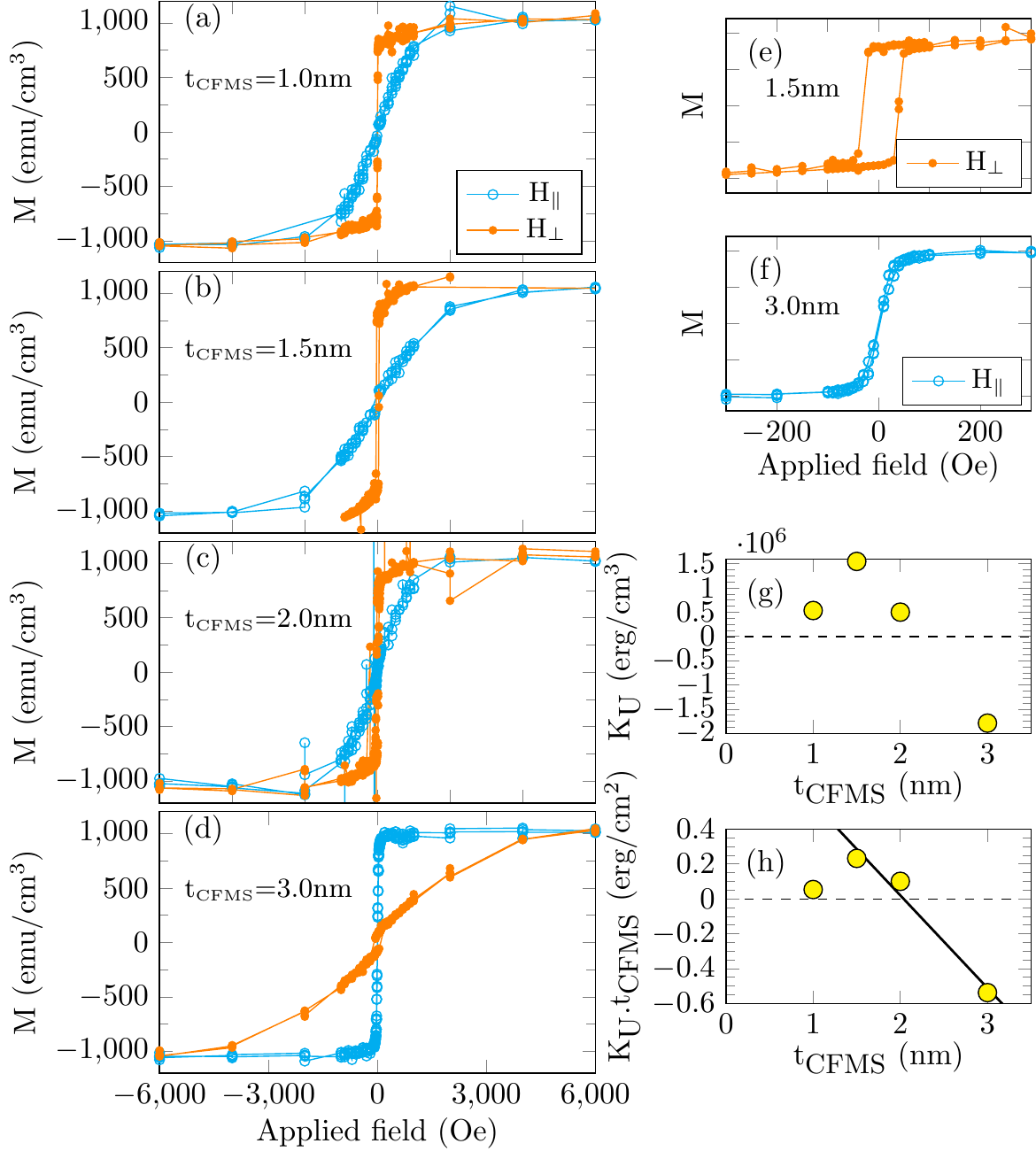}
\caption{\label{Fig:1} (a)-(d) SQUID magnetization hysteresis loops measured with the field in-plane (cyan/open symbols) and perpendicular-to-plane (orange/closed symbols) for MgO/CFMS/Pd trilayers. For t$_{\tiny\textrm{CFMS}}= 1, 1.5, 2~\textrm{nm}$ the samples show PMA with an out-of-plane easy axis of magnetization, while the t$_{\tiny\textrm{CFMS}}= 3~ \textrm{nm}$ sample in (d) has an in-plane easy axis. (e) The low field region shows the hysteresis and remanence for the out-of-plane measurement for t$_{\tiny\textrm{CFMS}}= 1.5~\textrm{nm}$. (f) The low field region of the t$_{\tiny\textrm{CFMS}}= 3~\textrm{nm}$ sample has a sharp change in magnetization with the field applied in-plane, although there is no hysteresis. (g) The uniaxial anisotropy is calculated from the data in panels (a)-(d). (h) Plotting K$_{\textrm{U}}$.t$_{\tiny\textrm{CFMS}}$ vs t$_{\tiny\textrm{CFMS}}$ shows the volume and surface contributions to the anisotropy. The solid line is a linear fit to the data excluding the t$_{\tiny\textrm{CFMS}}= 1~\textrm{nm}$ datapoint.}
\end{figure}

In Figs.\,1(a)-(d), magnetization measurements with the field applied parallel to the film plane and perpendicular to the film plane demonstrate PMA for CFMS films with thicknesses between 1 \& 2~nm. Samples in this thickness range are easily magnetized out-of-plane, having a small saturation field ($H_s < 100$ Oe), and high remanence [Fig.\,1(e)]. With the field applied in-plane, a larger applied field is required to saturate the magnetization, and the remanence is close to zero. Conversely, the magnetic behavior of a 3\,nm thick CFMS film shown in Fig.\,1(d) indicates an in-plane easy axis of magnetization.

The uniaxial magnetic anisotropy energy density $K_U$ is a quantitative measure of the PMA strength, and is determined from the difference in area under the out-of-plane and in-plane magnetization curves, where positive values correspond to PMA. $K_U$ is plotted as a function of t$_{\tiny\textrm{CFMS}}$ in Fig.\,\ref{Fig:1}(g), showing that the PMA is strongest for the t$_{\tiny\textrm{CFMS}} = 1.5$~nm film, having a value $K_U = 1.5\times 10^6$~erg/cm$^3$. This is comparable to values found for other Heusler alloys, which are typically $1-3\times 10^6$~erg/cm$^3$.\cite{Wang2010,Matsushita2015,Kamada2014,Wen2011} 

The decreasing $K_U$ with increasing film thickness and the transition to in-plane magnetic anisotropy (negative $K_U$) in Fig.\,1(g) is a feature of magnetic thin films usually attributed to the competition between interface induced PMA and the volume anisotropy, which tends to favor an in-plane easy axis.\cite{Vaz2008a} The uniaxial anisotropy is given by $K_U=K_V - 2 \pi M_s^2 + K_S/t$, where $K_V$ and $K_S$ are the bulk and interface anisotropy terms respectively, and the term $2 \pi M_s^2$ is the shape anisotropy.  $K_U.t_{\tiny\textrm{CFMS}}$ is plotted against t$_{\tiny\textrm{CFMS}}$ in Fig.\,\ref{Fig:1}(h), where the intercept of the linear extrapolation indicates $K_S=1.08$\,erg/cm$^2$. The slope is equal to the effective volume contribution $K_{V_{eff}} = K_V - 2\pi M_s^2 = 5.3\times 10^6$~erg/cm$^3$. The value of $K_S$ reported here is similar to that reported in Co$_2$FeAl\cite{Gabor2013,Wen2011} ($K_S=0.8-1.0$\,erg/cm$^2$), but larger than what has been measured in Co$_2$MnSi stacks\cite{Kamada2014} ($K_S=0.5$\,erg/cm$^2$) or multilayers\cite{Matsushita2015} ($K_S=0.16$\,erg/cm$^2$).

\begin{figure}[t!]
\includegraphics[width=8.5cm]{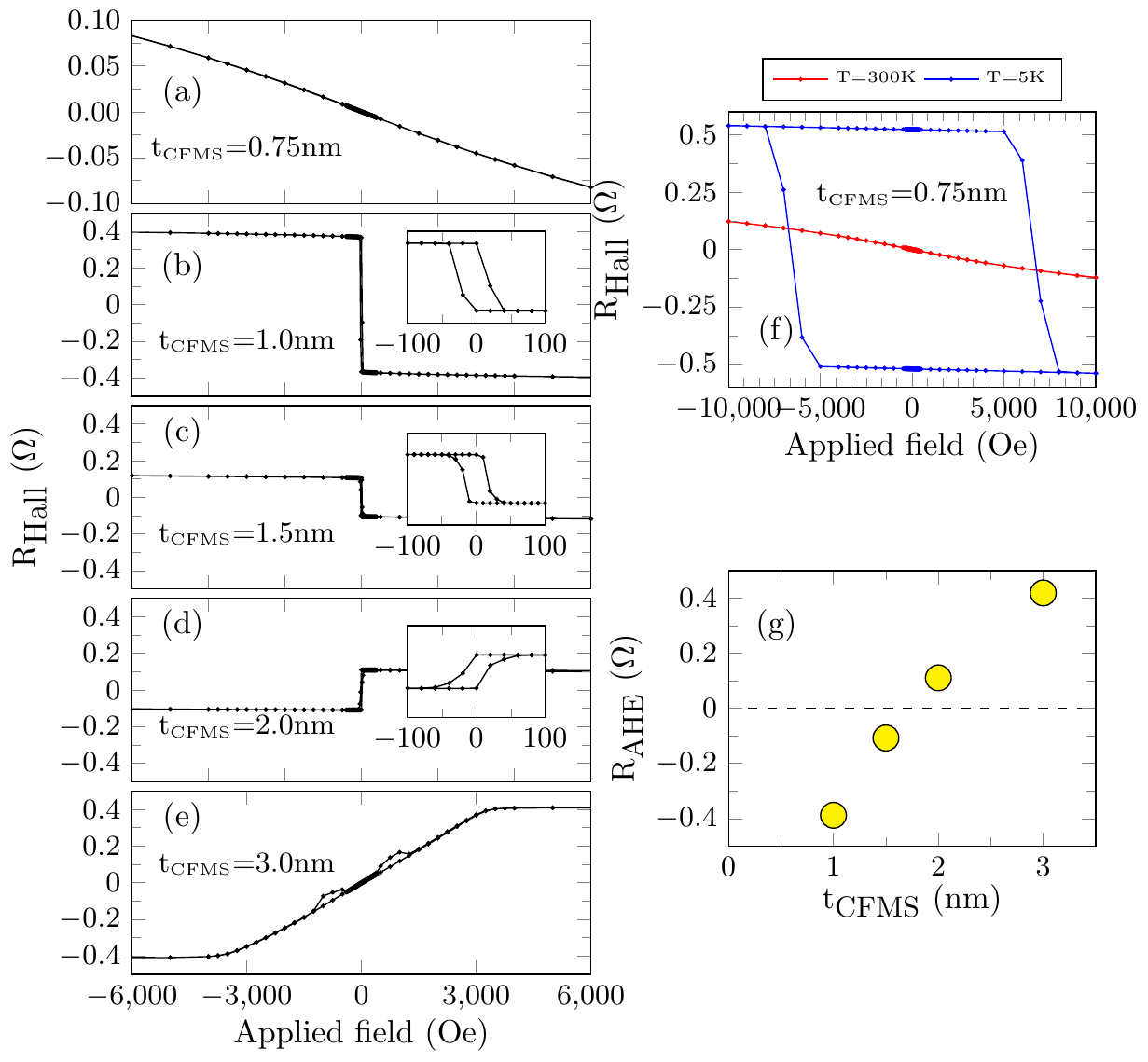}
\caption{\label{Fig:2} (a)-(e) Hall resistance of MgO/CFMS/Pd trilayers with $0.75 \leq t_{\tiny\textrm{CFMS}} \leq 3~\textrm{nm}$.  (f) Temperature dependent Hall measurements for the sample with t$_{\tiny\textrm{CFMS}}= 0.75~\textrm{nm}$ showing evidence of superparamagnetism. The AHE coefficient shown in (g) is determined by the zero-field extrapolation of the positive high-field Hall data. R$_{\tiny\textrm{AHE}}$ shows an approximately linear dependence on t$_{\tiny\textrm{CFMS}}$ with a sign change between 1.5 and 2~nm.}
\end{figure}


The Hall resistivity measured in a ferromagnetic material is empirically given by $\rho_{xy} = R_H H_z + R_{AHE} M_z$.\cite{Nagaosa2010} It is the sum of the normal Hall effect, linear in applied field ($H_z$), and the anomalous Hall effect (AHE), which is proportional to the out-of-plane magnetization ($M_z$). Therefore, measurements of the Hall effect can be used to probe PMA in thin films. The AHE in Figs.\,2(a)-(e) confirms the PMA for CFMS film thicknesses 1~nm $\leq$ t$_{\tiny \textrm{CFMS}}$ $\leq$ 2~nm, which show 100 \% remanence and a coercive field of around 25 Oe. The thicker 3~nm CFMS film in Fig.\,\ref{Fig:2}(e) shows an AHE characteristic of an in-plane easy magnetic axis, with a saturation field of about 3 kOe, in agreement with the magnetisation measurements in Fig.\,\ref{Fig:1}(d).

Data for the 0.75~nm CFMS film shown in Fig.\,2(a) shows superparamagnetic behaviour, similar to observations in CoFeB thin films below the thickness threshold for PMA.\cite{Zhu2014} A large, hysteretic AHE only becomes apparent at low temperature, as shown in Fig.\,2(f). This is probably an indication that the film is not continuous, with the discontinuous regions of the film acting as superparamagnetic particles at room temperature.



\begin{figure*}[t!]
	\includegraphics[width=0.8\linewidth]{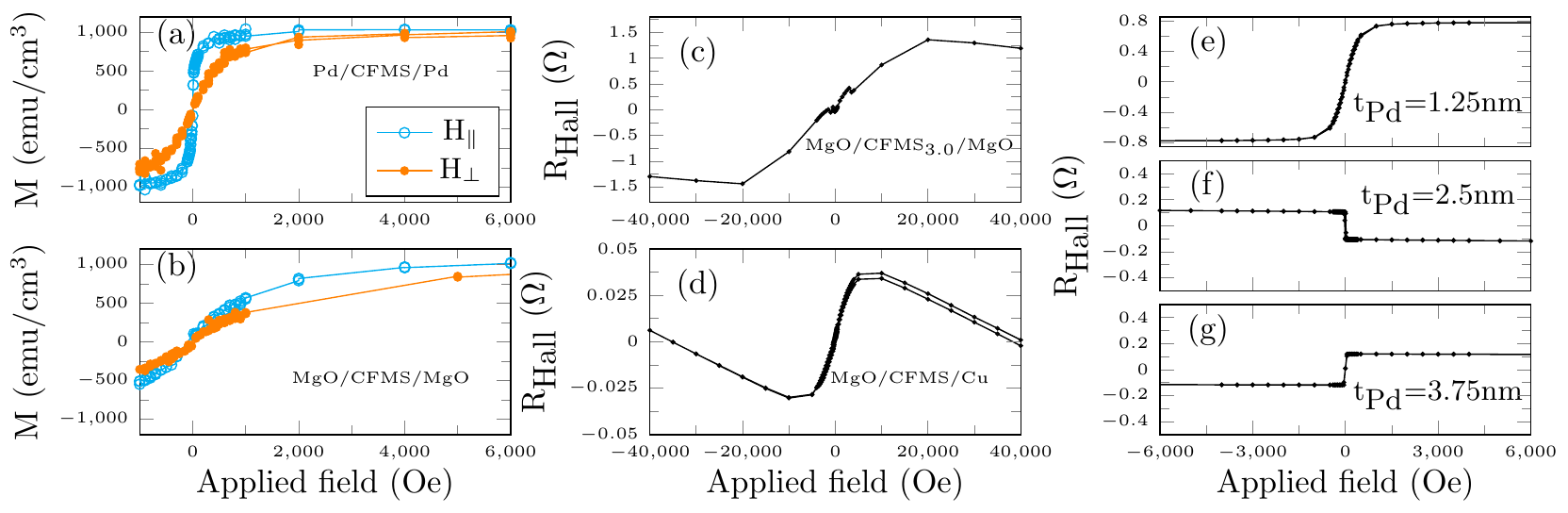}
	\caption{\label{Fig:3} (a) SQUID magnetization of a Pd/CFMS(1.5)/Pd trilayer shows in-plane magnetic anisotropy. Similarly, a MgO/CFMS(1.5)/MgO trilayer in (b) shows no evidence of PMA. (c) Hall measurements on a similar stack with a 3~nm CFMS layer show in-plane magnetic anisotropy, as expected. Replacing the Pd with Cu results in in-plane magnetic anisotropy, shown in the Hall measurement in (d). Hall measurements of MgO/CFMS/Pd trilayers with varying Pd thickness are shown in (e)-(g). A thin Pd layer (t$_{\tiny\textrm{Pd}} = 1.25~nm$) in (e) gives no PMA, while thicker Pd layers do, as shown by the sharp, hysteretic AHE observed for t$_{\tiny\textrm{Pd}} \geq 2.5 ~nm$ in (f) and (g). }
\end{figure*}

PMA in magnetic thin films generally results from a modification of the orbital angular momentum due to hybridization of orbitals at the interfaces. In MgO/CoFeB, it is thought to be the Fe-O hybridization at the interface that leads to the PMA\cite{Ikeda2010} although a thin metallic capping layer, often Ta, has been shown to contribute also.\cite{Lee2011,Liu2012b} PMA has also been observed and studied in Co/Pd and Co/Pt multilayers,\cite{Zeper1991} but in these cases, it is the Co $3d$-(Pd, Pt) $5d$ hybridization that is understood to induce the PMA.\cite{Weller1994,Nakajima1998} While the data presented here indicate an interfacial origin of the PMA in CFMS thin films, samples with different interfaces were prepared in order to understand which interface is important.

\textit{(i)}	Pd(2.5)/CFMS(1.5)/Pd(2.5): SQUID measurements show an in-plane easy axis for this stack [Fig.\,3(a)], demonstrating that the PMA is not from the CFMS-Pd interface alone, and that the MgO layer plays an important role.

\textit{(ii)}	MgO(2)/CFMS(1.5)/MgO(2): Magnetization measurements of this stack in Fig.\,3(b) show no evidence of PMA. Interestingly, Hall and resistivity measurements were not possible on this sample, suggesting the metallic CFMS layer does not form a continuous film. Hall measurements were made on a similar stack with a thicker CFMS layer (t$_{\tiny\textrm{CFMS}} = 3$~nm) and are shown in Fig.\,3(c). The AHE with a large saturation field is consistent with an in-plane easy axis.

\textit{(iii)} MgO(2)/CFMS(1.5)/Cu(3): Replacing the Pd capping layer with a Cu layer of similar thickness (3~nm) also destroyed the PMA. The Hall effect measurement in Fig.\,\ref{Fig:3}(d) shows a large saturation field of 5 kOe and zero remanence. This demonstrates the importance of the CFMS/Pd interface for attaining PMA. 

The thickness of the Pd layer also plays a role in attaining PMA in the CFMS thin film. Figs.\,\ref{Fig:3}(e)-(g) show Hall measurements for layers of the structure MgO(2)/CFMS(1.5)/Pd(t$_{\tiny\textrm{Pd}}$) for t$_{\tiny\textrm{Pd}}=1.25, 2.5, 3.75$~nm. For t$_{\tiny\textrm{Pd}}\geq 2.5$~nm, the CFMS has PMA, evidenced by the sharp, hysteretic AHE in Figs.\,3(f) and (g). For the thinnest Pd layer (t$_{\tiny\textrm{Pd}}=1.25$~nm), the film reverts to an in-plane magnetic anisotropy with a saturation field of about 1\,kOe and no remanence or coercivity. This is again similar to what is reported for CoFeB on MgO, where a Ta layer above a threshold thickness is required to attain PMA.\cite{Zhu2014a}


While the AHE is proportional to the out-of-plane magnetization, its sign and magnitude also depend on a number of other properties of the material, including the band structure, and various scattering mechanisms.\cite{Nagaosa2010} The AHE coefficient is defined here as the zero-field extrapolation of the high-field Hall effect data for positive applied fields. For CFMS thin films in a MgO/CFMS/Pd stack, we observe a thickness dependent AHE, as shown in Fig.\ref{Fig:2}(g). There is a change in the sign of R$_{\tiny\textrm{AHE}}$ between t$_{\tiny\textrm{CFMS}}=1.5$ and t$_{\tiny\textrm{CFMS}}=2.0$~nm. This trend, and the sign change, appears independent of the magnetic anisotropy, which transforms from PMA to in-plane anisotropy between t$_{\tiny\textrm{CFMS}}=2.0$ and t$_{\tiny\textrm{CFMS}}=3.0$~nm. It is worth noting that this AHE sign change also occurs with a change in the Pd capping layer thickness, shown in Figs.\,\ref{Fig:2}(e-g). While further work is required to understand the origin of this effect, we note that the AHE depends on spin- and spin-orbit dependent scattering,\cite{Nagaosa2010} and this system provides an opportunity to correlate these phenomena with the structural, magnetic and electronic properties of a highly spin-polarized magnetic material.


In conclusion, we have demonstrated PMA in Co$_2$Fe$_x$Mn$_{1-x}$Si with $x=0.4$ for thin films in a MgO/CFMS/Pd trilayer stack. Both the MgO and the Pd layers are necessary to generate the PMA. This stack shows promise for incorporation into a spin-transfer-torque device with high thermal stability, low switching current, and high output power. The observation of a thickness-tunable AHE is an interesting, novel effect, and further work is required to elucidate its origin.

The authors are grateful to Andrew Best from Callaghan Innovation for technical support, and to Sarah Spencer for the SEM measurements. B.M.L. acknowledges post-doctoral funding from the MacDiarmid Institute.  This work was supported by project funding from the MacDiarmid Institute and the New Zealand Ministry of Building, Innovation \& Employment Magnetic Devices contract RTVU1203. 


%
%

%


\bibliography{bib}

\begin{thebibliography}{24}%
\makeatletter
\providecommand \@ifxundefined [1]{%
 \@ifx{#1\undefined}
}%
\providecommand \@ifnum [1]{%
 \ifnum #1\expandafter \@firstoftwo
 \else \expandafter \@secondoftwo
 \fi
}%
\providecommand \@ifx [1]{%
 \ifx #1\expandafter \@firstoftwo
 \else \expandafter \@secondoftwo
 \fi
}%
\providecommand \natexlab [1]{#1}%
\providecommand \enquote  [1]{``#1''}%
\providecommand \bibnamefont  [1]{#1}%
\providecommand \bibfnamefont [1]{#1}%
\providecommand \citenamefont [1]{#1}%
\providecommand \href@noop [0]{\@secondoftwo}%
\providecommand \href [0]{\begingroup \@sanitize@url \@href}%
\providecommand \@href[1]{\@@startlink{#1}\@@href}%
\providecommand \@@href[1]{\endgroup#1\@@endlink}%
\providecommand \@sanitize@url [0]{\catcode `\\12\catcode `\$12\catcode
  `\&12\catcode `\#12\catcode `\^12\catcode `\_12\catcode `\%12\relax}%
\providecommand \@@startlink[1]{}%
\providecommand \@@endlink[0]{}%
\providecommand \url  [0]{\begingroup\@sanitize@url \@url }%
\providecommand \@url [1]{\endgroup\@href {#1}{\urlprefix }}%
\providecommand \urlprefix  [0]{URL }%
\providecommand \Eprint [0]{\href }%
\providecommand \doibase [0]{http://dx.doi.org/}%
\providecommand \selectlanguage [0]{\@gobble}%
\providecommand \bibinfo  [0]{\@secondoftwo}%
\providecommand \bibfield  [0]{\@secondoftwo}%
\providecommand \translation [1]{[#1]}%
\providecommand \BibitemOpen [0]{}%
\providecommand \bibitemStop [0]{}%
\providecommand \bibitemNoStop [0]{.\EOS\space}%
\providecommand \EOS [0]{\spacefactor3000\relax}%
\providecommand \BibitemShut  [1]{\csname bibitem#1\endcsname}%
\let\auto@bib@innerbib\@empty
\bibitem [{\citenamefont {Hirohata}\ \emph {et~al.}(2015)\citenamefont
  {Hirohata}, \citenamefont {Sukegawa}, \citenamefont {Yanagihara},
  \citenamefont {Zutic}, \citenamefont {Seki}, \citenamefont {Mizukami},\ and\
  \citenamefont {Swaminathan}}]{Hirohata2015}%
  \BibitemOpen
  \bibfield  {author} {\bibinfo {author} {\bibfnamefont {A.}~\bibnamefont
  {Hirohata}}, \bibinfo {author} {\bibfnamefont {H.}~\bibnamefont {Sukegawa}},
  \bibinfo {author} {\bibfnamefont {H.}~\bibnamefont {Yanagihara}}, \bibinfo
  {author} {\bibfnamefont {I.}~\bibnamefont {Zutic}}, \bibinfo {author}
  {\bibfnamefont {T.}~\bibnamefont {Seki}}, \bibinfo {author} {\bibfnamefont
  {S.}~\bibnamefont {Mizukami}}, \ and\ \bibinfo {author} {\bibfnamefont
  {R.}~\bibnamefont {Swaminathan}},\ }\href {\doibase
  10.1109/TMAG.2015.2457393} {\bibfield  {journal} {\bibinfo  {journal} {IEEE
  Trans. Magn.}\ }\textbf {\bibinfo {volume} {51}},\ \bibinfo {pages} {1}
  (\bibinfo {year} {2015})}\BibitemShut {NoStop}%
\bibitem [{\citenamefont {Mangin}\ \emph {et~al.}(2006)\citenamefont {Mangin},
  \citenamefont {Ravelosona}, \citenamefont {Katine}, \citenamefont {Carey},
  \citenamefont {Terris},\ and\ \citenamefont {Fullerton}}]{Mangin2006}%
  \BibitemOpen
  \bibfield  {author} {\bibinfo {author} {\bibfnamefont {S.}~\bibnamefont
  {Mangin}}, \bibinfo {author} {\bibfnamefont {D.}~\bibnamefont {Ravelosona}},
  \bibinfo {author} {\bibfnamefont {J.~A.}\ \bibnamefont {Katine}}, \bibinfo
  {author} {\bibfnamefont {M.~J.}\ \bibnamefont {Carey}}, \bibinfo {author}
  {\bibfnamefont {B.~D.}\ \bibnamefont {Terris}}, \ and\ \bibinfo {author}
  {\bibfnamefont {E.~E.}\ \bibnamefont {Fullerton}},\ }\href {\doibase
  10.1038/nmat1595} {\bibfield  {journal} {\bibinfo  {journal} {Nat. Mater.}\
  }\textbf {\bibinfo {volume} {5}},\ \bibinfo {pages} {210} (\bibinfo {year}
  {2006})}\BibitemShut {NoStop}%
\bibitem [{\citenamefont {Ikeda}\ \emph {et~al.}(2010)\citenamefont {Ikeda},
  \citenamefont {Miura}, \citenamefont {Yamamoto}, \citenamefont {Mizunuma},
  \citenamefont {Gan}, \citenamefont {Endo}, \citenamefont {Kanai},
  \citenamefont {Hayakawa}, \citenamefont {Matsukura},\ and\ \citenamefont
  {Ohno}}]{Ikeda2010}%
  \BibitemOpen
  \bibfield  {author} {\bibinfo {author} {\bibfnamefont {S.}~\bibnamefont
  {Ikeda}}, \bibinfo {author} {\bibfnamefont {K.}~\bibnamefont {Miura}},
  \bibinfo {author} {\bibfnamefont {H.}~\bibnamefont {Yamamoto}}, \bibinfo
  {author} {\bibfnamefont {K.}~\bibnamefont {Mizunuma}}, \bibinfo {author}
  {\bibfnamefont {H.~D.}\ \bibnamefont {Gan}}, \bibinfo {author} {\bibfnamefont
  {M.}~\bibnamefont {Endo}}, \bibinfo {author} {\bibfnamefont {S.}~\bibnamefont
  {Kanai}}, \bibinfo {author} {\bibfnamefont {J.}~\bibnamefont {Hayakawa}},
  \bibinfo {author} {\bibfnamefont {F.}~\bibnamefont {Matsukura}}, \ and\
  \bibinfo {author} {\bibfnamefont {H.}~\bibnamefont {Ohno}},\ }\href {\doibase
  10.1038/nmat2804} {\bibfield  {journal} {\bibinfo  {journal} {Nat. Mater.}\
  }\textbf {\bibinfo {volume} {9}},\ \bibinfo {pages} {721} (\bibinfo {year}
  {2010})}\BibitemShut {NoStop}%
\bibitem [{\citenamefont {Katsnelson}\ \emph {et~al.}(2008)\citenamefont
  {Katsnelson}, \citenamefont {Irkhin}, \citenamefont {Chioncel}, \citenamefont
  {Lichtenstein},\ and\ \citenamefont {de~Groot}}]{Katsnelson2008}%
  \BibitemOpen
  \bibfield  {author} {\bibinfo {author} {\bibfnamefont {M.~I.}\ \bibnamefont
  {Katsnelson}}, \bibinfo {author} {\bibfnamefont {V.~Y.}\ \bibnamefont
  {Irkhin}}, \bibinfo {author} {\bibfnamefont {L.}~\bibnamefont {Chioncel}},
  \bibinfo {author} {\bibfnamefont {A.~I.}\ \bibnamefont {Lichtenstein}}, \
  and\ \bibinfo {author} {\bibfnamefont {R.~A.}\ \bibnamefont {de~Groot}},\
  }\href {\doibase 10.1103/RevModPhys.80.315} {\bibfield  {journal} {\bibinfo
  {journal} {Rev. Mod. Phys.}\ }\textbf {\bibinfo {volume} {80}},\ \bibinfo
  {pages} {315} (\bibinfo {year} {2008})}\BibitemShut {NoStop}%
\bibitem [{\citenamefont {Kandpal}\ \emph {et~al.}(2006)\citenamefont
  {Kandpal}, \citenamefont {Fecher}, \citenamefont {Felser},\ and\
  \citenamefont {Sch{\"{o}}nhense}}]{Kandpal2006}%
  \BibitemOpen
  \bibfield  {author} {\bibinfo {author} {\bibfnamefont {H.~C.}\ \bibnamefont
  {Kandpal}}, \bibinfo {author} {\bibfnamefont {G.~H.}\ \bibnamefont {Fecher}},
  \bibinfo {author} {\bibfnamefont {C.}~\bibnamefont {Felser}}, \ and\ \bibinfo
  {author} {\bibfnamefont {G.}~\bibnamefont {Sch{\"{o}}nhense}},\ }\href
  {\doibase 10.1103/PhysRevB.73.094422} {\bibfield  {journal} {\bibinfo
  {journal} {Phys. Rev. B}\ }\textbf {\bibinfo {volume} {73}},\ \bibinfo
  {pages} {094422} (\bibinfo {year} {2006})}\BibitemShut {NoStop}%
\bibitem [{\citenamefont {Yang}, \citenamefont {Wei},\ and\ \citenamefont
  {Chen}(2013)}]{Yang2013}%
  \BibitemOpen
  \bibfield  {author} {\bibinfo {author} {\bibfnamefont {F.~J.}\ \bibnamefont
  {Yang}}, \bibinfo {author} {\bibfnamefont {C.}~\bibnamefont {Wei}}, \ and\
  \bibinfo {author} {\bibfnamefont {X.~Q.}\ \bibnamefont {Chen}},\ }\href
  {\doibase 10.1063/1.4803537} {\bibfield  {journal} {\bibinfo  {journal}
  {Appl. Phys. Lett.}\ }\textbf {\bibinfo {volume} {102}},\ \bibinfo {pages}
  {172403} (\bibinfo {year} {2013})}\BibitemShut {NoStop}%
\bibitem [{\citenamefont {Jourdan}\ \emph {et~al.}(2014)\citenamefont
  {Jourdan}, \citenamefont {Min{\'{a}}r}, \citenamefont {Braun}, \citenamefont
  {Kronenberg}, \citenamefont {Chadov}, \citenamefont {Balke}, \citenamefont
  {Gloskovskii}, \citenamefont {Kolbe}, \citenamefont {Elmers}, \citenamefont
  {Sch{\"{o}}nhense}, \citenamefont {Ebert}, \citenamefont {Felser},\ and\
  \citenamefont {Kl{\"{a}}ui}}]{Jourdan2014}%
  \BibitemOpen
  \bibfield  {author} {\bibinfo {author} {\bibfnamefont {M.}~\bibnamefont
  {Jourdan}}, \bibinfo {author} {\bibfnamefont {J.}~\bibnamefont
  {Min{\'{a}}r}}, \bibinfo {author} {\bibfnamefont {J.}~\bibnamefont {Braun}},
  \bibinfo {author} {\bibfnamefont {A.}~\bibnamefont {Kronenberg}}, \bibinfo
  {author} {\bibfnamefont {S.}~\bibnamefont {Chadov}}, \bibinfo {author}
  {\bibfnamefont {B.}~\bibnamefont {Balke}}, \bibinfo {author} {\bibfnamefont
  {A.}~\bibnamefont {Gloskovskii}}, \bibinfo {author} {\bibfnamefont
  {M.}~\bibnamefont {Kolbe}}, \bibinfo {author} {\bibfnamefont
  {H.}~\bibnamefont {Elmers}}, \bibinfo {author} {\bibfnamefont
  {G.}~\bibnamefont {Sch{\"{o}}nhense}}, \bibinfo {author} {\bibfnamefont
  {H.}~\bibnamefont {Ebert}}, \bibinfo {author} {\bibfnamefont
  {C.}~\bibnamefont {Felser}}, \ and\ \bibinfo {author} {\bibfnamefont
  {M.}~\bibnamefont {Kl{\"{a}}ui}},\ }\href {\doibase 10.1038/ncomms4974}
  {\bibfield  {journal} {\bibinfo  {journal} {Nat. Commun.}\ }\textbf {\bibinfo
  {volume} {5}},\ \bibinfo {pages} {3974} (\bibinfo {year} {2014})}\BibitemShut
  {NoStop}%
\bibitem [{\citenamefont {Kubota}\ \emph {et~al.}(2009)\citenamefont {Kubota},
  \citenamefont {Tsunegi}, \citenamefont {Oogane}, \citenamefont {Mizukami},
  \citenamefont {Miyazaki}, \citenamefont {Naganuma},\ and\ \citenamefont
  {Ando}}]{Kubota2009}%
  \BibitemOpen
  \bibfield  {author} {\bibinfo {author} {\bibfnamefont {T.}~\bibnamefont
  {Kubota}}, \bibinfo {author} {\bibfnamefont {S.}~\bibnamefont {Tsunegi}},
  \bibinfo {author} {\bibfnamefont {M.}~\bibnamefont {Oogane}}, \bibinfo
  {author} {\bibfnamefont {S.}~\bibnamefont {Mizukami}}, \bibinfo {author}
  {\bibfnamefont {T.}~\bibnamefont {Miyazaki}}, \bibinfo {author}
  {\bibfnamefont {H.}~\bibnamefont {Naganuma}}, \ and\ \bibinfo {author}
  {\bibfnamefont {Y.}~\bibnamefont {Ando}},\ }\href {\doibase
  10.1063/1.3105982} {\bibfield  {journal} {\bibinfo  {journal} {Appl. Phys.
  Lett.}\ }\textbf {\bibinfo {volume} {94}},\ \bibinfo {pages} {122504}
  (\bibinfo {year} {2009})}\BibitemShut {NoStop}%
\bibitem [{\citenamefont {Sato}\ \emph {et~al.}(2011)\citenamefont {Sato},
  \citenamefont {Oogane}, \citenamefont {Naganuma},\ and\ \citenamefont
  {Ando}}]{Sato2011}%
  \BibitemOpen
  \bibfield  {author} {\bibinfo {author} {\bibfnamefont {J.}~\bibnamefont
  {Sato}}, \bibinfo {author} {\bibfnamefont {M.}~\bibnamefont {Oogane}},
  \bibinfo {author} {\bibfnamefont {H.}~\bibnamefont {Naganuma}}, \ and\
  \bibinfo {author} {\bibfnamefont {Y.}~\bibnamefont {Ando}},\ }\href {\doibase
  10.1143/APEX.4.113005} {\bibfield  {journal} {\bibinfo  {journal} {Appl.
  Phys. Express}\ }\textbf {\bibinfo {volume} {4}},\ \bibinfo {pages} {113005}
  (\bibinfo {year} {2011})}\BibitemShut {NoStop}%
\bibitem [{\citenamefont {Wen}\ \emph {et~al.}(2011)\citenamefont {Wen},
  \citenamefont {Sukegawa}, \citenamefont {Mitani},\ and\ \citenamefont
  {Inomata}}]{Wen2011}%
  \BibitemOpen
  \bibfield  {author} {\bibinfo {author} {\bibfnamefont {Z.}~\bibnamefont
  {Wen}}, \bibinfo {author} {\bibfnamefont {H.}~\bibnamefont {Sukegawa}},
  \bibinfo {author} {\bibfnamefont {S.}~\bibnamefont {Mitani}}, \ and\ \bibinfo
  {author} {\bibfnamefont {K.}~\bibnamefont {Inomata}},\ }\href {\doibase
  10.1063/1.3600645} {\bibfield  {journal} {\bibinfo  {journal} {Appl. Phys.
  Lett.}\ }\textbf {\bibinfo {volume} {98}},\ \bibinfo {pages} {2011} (\bibinfo
  {year} {2011})}\BibitemShut {NoStop}%
\bibitem [{\citenamefont {Li}\ \emph {et~al.}(2011)\citenamefont {Li},
  \citenamefont {Yin}, \citenamefont {Liu}, \citenamefont {Zhang},
  \citenamefont {Xu}, \citenamefont {Miao},\ and\ \citenamefont
  {Jiang}}]{Li2011a}%
  \BibitemOpen
  \bibfield  {author} {\bibinfo {author} {\bibfnamefont {X.}~\bibnamefont
  {Li}}, \bibinfo {author} {\bibfnamefont {S.}~\bibnamefont {Yin}}, \bibinfo
  {author} {\bibfnamefont {Y.}~\bibnamefont {Liu}}, \bibinfo {author}
  {\bibfnamefont {D.}~\bibnamefont {Zhang}}, \bibinfo {author} {\bibfnamefont
  {X.}~\bibnamefont {Xu}}, \bibinfo {author} {\bibfnamefont {J.}~\bibnamefont
  {Miao}}, \ and\ \bibinfo {author} {\bibfnamefont {Y.}~\bibnamefont {Jiang}},\
  }\href {\doibase 10.1143/APEX.4.043006} {\bibfield  {journal} {\bibinfo
  {journal} {Appl. Phys. Express}\ }\textbf {\bibinfo {volume} {4}},\ \bibinfo
  {pages} {043006} (\bibinfo {year} {2011})}\BibitemShut {NoStop}%
\bibitem [{\citenamefont {Matsushita}\ \emph {et~al.}(2015)\citenamefont
  {Matsushita}, \citenamefont {Takamura}, \citenamefont {Fujino}, \citenamefont
  {Sonobe},\ and\ \citenamefont {Nakagawa}}]{Matsushita2015}%
  \BibitemOpen
  \bibfield  {author} {\bibinfo {author} {\bibfnamefont {N.}~\bibnamefont
  {Matsushita}}, \bibinfo {author} {\bibfnamefont {Y.}~\bibnamefont
  {Takamura}}, \bibinfo {author} {\bibfnamefont {Y.}~\bibnamefont {Fujino}},
  \bibinfo {author} {\bibfnamefont {Y.}~\bibnamefont {Sonobe}}, \ and\ \bibinfo
  {author} {\bibfnamefont {S.}~\bibnamefont {Nakagawa}},\ }\href {\doibase
  10.1063/1.4907892} {\bibfield  {journal} {\bibinfo  {journal} {Appl. Phys.
  Lett.}\ }\textbf {\bibinfo {volume} {106}},\ \bibinfo {pages} {062403}
  (\bibinfo {year} {2015})}\BibitemShut {NoStop}%
\bibitem [{\citenamefont {Kamada}\ \emph {et~al.}(2014)\citenamefont {Kamada},
  \citenamefont {Kubota}, \citenamefont {Takahashi}, \citenamefont {Sonobe},\
  and\ \citenamefont {Takanashi}}]{Kamada2014}%
  \BibitemOpen
  \bibfield  {author} {\bibinfo {author} {\bibfnamefont {T.}~\bibnamefont
  {Kamada}}, \bibinfo {author} {\bibfnamefont {T.}~\bibnamefont {Kubota}},
  \bibinfo {author} {\bibfnamefont {S.}~\bibnamefont {Takahashi}}, \bibinfo
  {author} {\bibfnamefont {Y.}~\bibnamefont {Sonobe}}, \ and\ \bibinfo {author}
  {\bibfnamefont {K.}~\bibnamefont {Takanashi}},\ }\href {\doibase
  10.1109/TMAG.2014.2322857} {\bibfield  {journal} {\bibinfo  {journal} {IEEE
  Trans. Magn.}\ }\textbf {\bibinfo {volume} {50}},\ \bibinfo {pages} {1}
  (\bibinfo {year} {2014})}\BibitemShut {NoStop}%
\bibitem [{\citenamefont {Wang}, \citenamefont {Sukegawa},\ and\ \citenamefont
  {Inomata}(2010)}]{Wang2010}%
  \BibitemOpen
  \bibfield  {author} {\bibinfo {author} {\bibfnamefont {W.}~\bibnamefont
  {Wang}}, \bibinfo {author} {\bibfnamefont {H.}~\bibnamefont {Sukegawa}}, \
  and\ \bibinfo {author} {\bibfnamefont {K.}~\bibnamefont {Inomata}},\ }\href
  {\doibase 10.1143/APEX.3.093002} {\bibfield  {journal} {\bibinfo  {journal}
  {Appl. Phys. Express}\ }\textbf {\bibinfo {volume} {3}},\ \bibinfo {pages}
  {093002} (\bibinfo {year} {2010})}\BibitemShut {NoStop}%
\bibitem [{\citenamefont {Vaz}, \citenamefont {Bland},\ and\ \citenamefont
  {Lauhoff}(2008)}]{Vaz2008a}%
  \BibitemOpen
  \bibfield  {author} {\bibinfo {author} {\bibfnamefont {C.~A.~F.}\
  \bibnamefont {Vaz}}, \bibinfo {author} {\bibfnamefont {J.~A.~C.}\
  \bibnamefont {Bland}}, \ and\ \bibinfo {author} {\bibfnamefont
  {G.}~\bibnamefont {Lauhoff}},\ }\href {\doibase
  10.1088/0034-4885/71/5/056501} {\bibfield  {journal} {\bibinfo  {journal}
  {Reports Prog. Phys.}\ }\textbf {\bibinfo {volume} {71}},\ \bibinfo {pages}
  {056501} (\bibinfo {year} {2008})}\BibitemShut {NoStop}%
\bibitem [{\citenamefont {Gabor}, \citenamefont {Petrisor},\ and\ \citenamefont
  {Tiusan}(2013)}]{Gabor2013}%
  \BibitemOpen
  \bibfield  {author} {\bibinfo {author} {\bibfnamefont {M.~S.}\ \bibnamefont
  {Gabor}}, \bibinfo {author} {\bibfnamefont {T.}~\bibnamefont {Petrisor}}, \
  and\ \bibinfo {author} {\bibfnamefont {C.}~\bibnamefont {Tiusan}},\ }\href
  {\doibase 10.1063/1.4818326} {\bibfield  {journal} {\bibinfo  {journal} {J.
  Appl. Phys.}\ }\textbf {\bibinfo {volume} {114}},\ \bibinfo {pages} {063905}
  (\bibinfo {year} {2013})}\BibitemShut {NoStop}%
\bibitem [{\citenamefont {Nagaosa}\ \emph {et~al.}(2010)\citenamefont
  {Nagaosa}, \citenamefont {Sinova}, \citenamefont {Onoda}, \citenamefont
  {MacDonald},\ and\ \citenamefont {Ong}}]{Nagaosa2010}%
  \BibitemOpen
  \bibfield  {author} {\bibinfo {author} {\bibfnamefont {N.}~\bibnamefont
  {Nagaosa}}, \bibinfo {author} {\bibfnamefont {J.}~\bibnamefont {Sinova}},
  \bibinfo {author} {\bibfnamefont {S.}~\bibnamefont {Onoda}}, \bibinfo
  {author} {\bibfnamefont {A.~H.}\ \bibnamefont {MacDonald}}, \ and\ \bibinfo
  {author} {\bibfnamefont {N.~P.}\ \bibnamefont {Ong}},\ }\href {\doibase
  10.1103/RevModPhys.82.1539} {\bibfield  {journal} {\bibinfo  {journal} {Rev.
  Mod. Phys.}\ }\textbf {\bibinfo {volume} {82}},\ \bibinfo {pages} {1539}
  (\bibinfo {year} {2010})}\BibitemShut {NoStop}%
\bibitem [{\citenamefont {Zhu}\ \emph {et~al.}(2014)\citenamefont {Zhu},
  \citenamefont {Chen}, \citenamefont {Zhang}, \citenamefont {Yu},\ and\
  \citenamefont {Liu}}]{Zhu2014}%
  \BibitemOpen
  \bibfield  {author} {\bibinfo {author} {\bibfnamefont {T.}~\bibnamefont
  {Zhu}}, \bibinfo {author} {\bibfnamefont {P.}~\bibnamefont {Chen}}, \bibinfo
  {author} {\bibfnamefont {Q.~H.}\ \bibnamefont {Zhang}}, \bibinfo {author}
  {\bibfnamefont {R.~C.}\ \bibnamefont {Yu}}, \ and\ \bibinfo {author}
  {\bibfnamefont {B.~G.}\ \bibnamefont {Liu}},\ }\href {\doibase
  10.1063/1.4878538} {\bibfield  {journal} {\bibinfo  {journal} {Appl. Phys.
  Lett.}\ }\textbf {\bibinfo {volume} {104}},\ \bibinfo {pages} {202404}
  (\bibinfo {year} {2014})}\BibitemShut {NoStop}%
\bibitem [{\citenamefont {Lee}\ \emph {et~al.}(2011)\citenamefont {Lee},
  \citenamefont {Sapan}, \citenamefont {Kang},\ and\ \citenamefont
  {Fullerton}}]{Lee2011}%
  \BibitemOpen
  \bibfield  {author} {\bibinfo {author} {\bibfnamefont {K.}~\bibnamefont
  {Lee}}, \bibinfo {author} {\bibfnamefont {J.~J.}\ \bibnamefont {Sapan}},
  \bibinfo {author} {\bibfnamefont {S.~H.}\ \bibnamefont {Kang}}, \ and\
  \bibinfo {author} {\bibfnamefont {E.~E.}\ \bibnamefont {Fullerton}},\ }\href
  {\doibase 10.1063/1.3592986} {\bibfield  {journal} {\bibinfo  {journal} {J.
  Appl. Phys.}\ }\textbf {\bibinfo {volume} {109}},\ \bibinfo {pages} {123910}
  (\bibinfo {year} {2011})}\BibitemShut {NoStop}%
\bibitem [{\citenamefont {Liu}, \citenamefont {Cai},\ and\ \citenamefont
  {Sun}(2012)}]{Liu2012b}%
  \BibitemOpen
  \bibfield  {author} {\bibinfo {author} {\bibfnamefont {T.}~\bibnamefont
  {Liu}}, \bibinfo {author} {\bibfnamefont {J.~W.}\ \bibnamefont {Cai}}, \ and\
  \bibinfo {author} {\bibfnamefont {L.}~\bibnamefont {Sun}},\ }\href {\doibase
  10.1063/1.4748337} {\bibfield  {journal} {\bibinfo  {journal} {AIP Adv.}\
  }\textbf {\bibinfo {volume} {2}},\ \bibinfo {pages} {032151} (\bibinfo {year}
  {2012})}\BibitemShut {NoStop}%
\bibitem [{\citenamefont {Zeper}\ \emph {et~al.}(1991)\citenamefont {Zeper},
  \citenamefont {van Kesteren}, \citenamefont {Jacobs}, \citenamefont
  {Spruit},\ and\ \citenamefont {Carcia}}]{Zeper1991}%
  \BibitemOpen
  \bibfield  {author} {\bibinfo {author} {\bibfnamefont {W.~B.}\ \bibnamefont
  {Zeper}}, \bibinfo {author} {\bibfnamefont {H.~W.}\ \bibnamefont {van
  Kesteren}}, \bibinfo {author} {\bibfnamefont {B.~A.~J.}\ \bibnamefont
  {Jacobs}}, \bibinfo {author} {\bibfnamefont {J.~H.~M.}\ \bibnamefont
  {Spruit}}, \ and\ \bibinfo {author} {\bibfnamefont {P.~F.}\ \bibnamefont
  {Carcia}},\ }\href {\doibase 10.1063/1.349419} {\bibfield  {journal}
  {\bibinfo  {journal} {J. Appl. Phys.}\ }\textbf {\bibinfo {volume} {70}},\
  \bibinfo {pages} {2264} (\bibinfo {year} {1991})}\BibitemShut {NoStop}%
\bibitem [{\citenamefont {Weller}\ \emph {et~al.}(1994)\citenamefont {Weller},
  \citenamefont {Wu}, \citenamefont {St{\"{o}}hr}, \citenamefont {Samant},
  \citenamefont {Hermsmeier},\ and\ \citenamefont {Chappert}}]{Weller1994}%
  \BibitemOpen
  \bibfield  {author} {\bibinfo {author} {\bibfnamefont {D.}~\bibnamefont
  {Weller}}, \bibinfo {author} {\bibfnamefont {Y.}~\bibnamefont {Wu}}, \bibinfo
  {author} {\bibfnamefont {J.}~\bibnamefont {St{\"{o}}hr}}, \bibinfo {author}
  {\bibfnamefont {M.~G.}\ \bibnamefont {Samant}}, \bibinfo {author}
  {\bibfnamefont {B.~D.}\ \bibnamefont {Hermsmeier}}, \ and\ \bibinfo {author}
  {\bibfnamefont {C.}~\bibnamefont {Chappert}},\ }\href {\doibase
  10.1103/PhysRevB.49.12888} {\bibfield  {journal} {\bibinfo  {journal} {Phys.
  Rev. B}\ }\textbf {\bibinfo {volume} {49}},\ \bibinfo {pages} {12888}
  (\bibinfo {year} {1994})}\BibitemShut {NoStop}%
\bibitem [{\citenamefont {Nakajima}\ \emph {et~al.}(1998)\citenamefont
  {Nakajima}, \citenamefont {Koide}, \citenamefont {Shidara}, \citenamefont
  {Miyauchi}, \citenamefont {Fukutani}, \citenamefont {Fujimori}, \citenamefont
  {Iio}, \citenamefont {Katayama}, \citenamefont {N{\'{y}}vlt},\ and\
  \citenamefont {Suzuki}}]{Nakajima1998}%
  \BibitemOpen
  \bibfield  {author} {\bibinfo {author} {\bibfnamefont {N.}~\bibnamefont
  {Nakajima}}, \bibinfo {author} {\bibfnamefont {T.}~\bibnamefont {Koide}},
  \bibinfo {author} {\bibfnamefont {T.}~\bibnamefont {Shidara}}, \bibinfo
  {author} {\bibfnamefont {H.}~\bibnamefont {Miyauchi}}, \bibinfo {author}
  {\bibfnamefont {H.}~\bibnamefont {Fukutani}}, \bibinfo {author}
  {\bibfnamefont {A.}~\bibnamefont {Fujimori}}, \bibinfo {author}
  {\bibfnamefont {K.}~\bibnamefont {Iio}}, \bibinfo {author} {\bibfnamefont
  {T.}~\bibnamefont {Katayama}}, \bibinfo {author} {\bibfnamefont
  {M.}~\bibnamefont {N{\'{y}}vlt}}, \ and\ \bibinfo {author} {\bibfnamefont
  {Y.}~\bibnamefont {Suzuki}},\ }\href {\doibase 10.1103/PhysRevLett.81.5229}
  {\bibfield  {journal} {\bibinfo  {journal} {Phys. Rev. Lett.}\ }\textbf
  {\bibinfo {volume} {81}},\ \bibinfo {pages} {5229} (\bibinfo {year}
  {1998})}\BibitemShut {NoStop}%
\bibitem [{\citenamefont {Zhu}(2014)}]{Zhu2014a}%
  \BibitemOpen
  \bibfield  {author} {\bibinfo {author} {\bibfnamefont {T.}~\bibnamefont
  {Zhu}},\ }\href {\doibase 10.1088/1674-1056/23/4/047504} {\bibfield
  {journal} {\bibinfo  {journal} {Chinese Phys. B}\ }\textbf {\bibinfo {volume}
  {23}},\ \bibinfo {pages} {047504} (\bibinfo {year} {2014})}\BibitemShut
  {NoStop}%
\end{thebibliography}%

\end{document}